# Probable projectile-target combinations for the synthesis of super heavy nucleus $^{286}112$

K. P. Santhosh and V. Bobby Jose

*School of Pure and Applied Physics, Kannur University, Swami Anandatheertha Campus, Payyanur 670327, Kerala, India.*

Email: drkpsanthosh@gmail.com

**Abstract**

The fusion cross sections for the reactions of all the projectile-target combinations found in the cold valleys of $^{286}112$ have been studied using scattering potential as the sum of Coulomb and proximity potential, so as to predict the most probable projectile-target combinations in heavy ion fusion reactions for the synthesis of super heavy nucleus $^{286}112$. While considering the nature of potential pockets and half lives of the colliding nuclei, the systems $^{82}$Ge + $^{204}$Hg, $^{80}$Ge + $^{206}$Hg and $^{78}$Zn + $^{208}$Pb found in the deep cold valley region and the systems $^{48}$Ca+$^{238}$U, $^{38}$S+$^{248}$Cm and $^{44}$Ar+$^{242}$Pu in the cold valleys are predicted to be the better optimal projectile-target combinations for the synthesis of super heavy nucleus $^{286}112$.

**Keywords:** Heavy-ion reactions; Coulomb and proximity potential

**PACS Nos.** 25.60.Pj; 25.70.Jj

## 1. Introduction

Extensive studies have been made both experimentally and theoretically on heavy-ion fusion reactions for the understanding of the reaction mechanisms involved, especially for the synthesis of super heavy elements (SHE), which is a hot topic and very interesting problem in Nuclear Physics. As the fusion-evaporation reactions give low cross sections, the increasing difficulty with which heavier SHE can be produced is a major challenge to experimental investigations, nowadays.

Followed by the discovery of radioactivity by Henri Becquerel in 1896 [1] thirty-one radioactive elements including U and Th, were added to the table of elements in the first century of the development of nuclear physics. In a retrospect, both theoretical and experimental, different periods have to be considered historically [2, 3] in reviewing a century of radioactive elements up to Z=119 [4]. Initially, a first period (1896–1939) [5] yielded the radioactive elements between Bi and U, where as second period (1934–1955) [6] produced large quantities of new manmade elements with the enormous developments in fission techniques and n, p, d, α-capture of the heaviest isotopes in the high neutron fluxes of nuclear reactors. The development of particle accelerators and particle detectors in the third period (1955–1974) [7], allowed scientists to fuse light elements B to O with long-lived isotopes of the heaviest actinides produced in nuclear reactors. In this production method, as the compound nuclei formed after fusion, are heated owing to excitation energies between 40 and 50 MeV, the method is called "hot fusion" or "actinide-based fusion." In the fourth period, which began in 1974 [8, 9] closed-shell nuclei, $^{208}$Pb and $^{209}$Bi, are fused with medium-weight neutron-rich isotopes such as $^{54}$Cr to $^{70}$Zn produced the elements 107 to 112 at GSI, Darmstadt [10–14]. In all the reactions, at excitation energies of 10–20 MeV the compound systems stay colder than in hot fusion reactions and the method was called "cold fusion" or more appropriately "cluster-based fusion" or the neutral "Pb/Bi-based fusion." Moreover, SHN with $Z$ =113–116 and 118 have been synthesized at JINR-FLNR, Dubna, in collaboration with the LLNL researchers [15-22] and very recently they were also successful in the synthesis of two isotopes of $Z$ =117.

The recent progress in the accelerator technologies has encouraged the experimentalist to reach the shore of the "magic island" or the *island of stability* up to around $Z$ = 120, 124, or 126 and $N$ =184 [23]. In hot fusion reactions with actinide targets such as $^{233,238}$U, $^{237}$Np, $^{242,244}$Pu, $^{243}$Am, $^{245,248}$Cm, and $^{249}$Cf and cold fusion reactions with Pb and Bi targets, a common limitation is the available choice of target and projectile combination for the synthesis of super heavy nuclei. In all the cases, due to the double magicity of $^{48}$Ca, similar with $^{208}$Pb, $^{48}$Ca was proposed [19, 22] as the projectile on various trans-uranium targets. The synthesis of many super heavy elements with $Z$ < 119, during last three decades is mainly based on this idea [21, 24]. Recently, Oganessian *et al.* have reported the synthesis of element 117 via the fusion of $^{48}$Ca and $^{249}$Bk [25, 26].

The study of super heavy elements leads to many new findings, especially the possible appearance of new magic shell numbers or more precisely the prediction of the doubly-magic nucleus next to $Z$ = 82, $N$ = 126, $^{208}$Pb. In addition to general radioactive decay through alpha and beta decay with subsequent emission of gamma rays, decay through spontaneous fission and cluster radioactivity [27] was predicted in recent years. Cluster radioactivity is the spontaneous decay of nuclei by the emission of particles heavier than alpha particle say $^{14}$C, $^{24}$Ne, $^{30}$Mg and

$^{34}$Si and therefore occupies intermediate position between alpha decay and spontaneous fission. Based on the in depth and wide theoretical [28-31] and experimental [32-36] studies on cluster radioactivity, it has been established that cluster decay is one of the key decay mode for the radioactive nuclei, especially in the super heavy region.

Sticking on the concept of cold valleys which were introduced in relation to the structure of minima in the so-called driving potential, which is the difference between the interaction potential and the decay energy Q of the reaction, radioactive decay of super heavy nuclei $^{286}$112, $^{292}$114, and $^{296}$116 were studied [37], using the Coulomb and Proximity Potential model (CPPM) [38] and it was found that in addition to alpha particle, $^{8}$Be, $^{14}$C, $^{28}$Mg, $^{34}$Si, $^{50}$Ca, etc. are optimal cases of cluster radioactivity, since they lie in the cold valleys. Two other regions of deep minima centered on $^{208}$Pb and $^{132}$Sn were also found. On the basis of the observation of the formation of excited compound nuclei $^{286}$112, $^{292}$114, and $^{296}$116 obtained during the fusion processes with $^{48}$Ca beam on $^{238}$U, $^{244}$Pu, and $^{248}$Cm at the same excitation energy $E^* = 33$ MeV [39], in the present work we have studied the fusion cross sections for the reactions of all the projectile-target combinations found in the cold valleys of $^{286}$112 [37], using scattering potential as the sum of the Coulomb and proximity potential, so as to predict the most probable projectile-target combinations in heavy ion fusion reactions for the synthesis of super heavy nuclei.

In the analysis of heavy-ion fusion reactions [40-49], an internuclear interaction consisting of repulsive Coulomb and centrifugal potentials and attractive nuclear potential plays a major role, where the potential is a function of the distance between centres-of mass of the colliding nuclei. At a distance referred to as Coulomb barrier the total potential attains a maximum value, where the repulsive and attractive forces balance each other and the energy of relative motion must overcome this barrier in order for the nuclei to be captured and fused.

## 2. Theory

### 2.1. The potential

Exploring the different nuclear reaction mechanisms, which are exclusively governed by the nucleus-nucleus potential, with a unique nuclear potential is an extensive challenge for the last several years in nuclear physics.

Explaining the nuclear potential as the product of a geometrical factor, which is proportional to the reduced radii of colliding nuclei and a universal function is commonly accepted, as it is incorporating the role of different colliding nuclei in the geometrical factor. In this effort, a simple formula for the nucleus-nucleus interaction energy as a function of separation between the surfaces of the approaching nuclei has been given by the proximity potential of Blocki et al. [50]. The formula is free of adjustable parameters and makes use of the measured values of the nuclear surface tension and surface diffuseness.

The interaction barrier for two colliding nuclei is given as:

$$V = \frac{Z_1 Z_2 e^2}{r} + V_P(z) + \frac{\hbar^2 \ell(\ell+1)}{2\mu r^2} \qquad (1)$$

where $Z_1$ and $Z_2$ are the atomic numbers of projectile and target, r is the distance between the centers of the projectile and target, z is the distance between the near surfaces of the projectile and target, $\ell$ is the angular momentum, $\mu$ is the reduced mass of the target and projectile and $V_P(z)$ is the proximity potential given as:

$$V_P(z) = 4\pi\gamma b \frac{C_1 C_2}{C_1 + C_2} \phi(\frac{z}{b}) \qquad (2)$$

with the nuclear surface tension coefficient,

$$\gamma = 0.9517[1 - 1.7826(N-Z)^2 / A^2] \qquad (3)$$

$\phi$, the universal proximity potential is given as:

$$\phi(\xi) = -4.41\exp(-\xi / 0.7176), \quad \text{for } \xi \geq 1.9475 \qquad (4)$$

$$\phi(\xi) = -1.7817 + 0.9270\xi + 0.01696\xi^2 - 0.05148\xi^3, \quad \text{for } 0 \leq \xi \leq 1.9475 \qquad (5)$$

$$\phi(\xi) = -1.7817 + 0.9270\xi + 0.0143\xi^2 - 0.09\xi^3, \quad \text{for } \xi \leq 0 \qquad (6)$$

with $\xi = z/b$, where the width (diffuseness) of nuclear surface $b \approx 1$ and Siissmann Central radii $C_i$ related to sharp radii $R_i$ as $C_i = R_i - \frac{b^2}{R_i}$. For $R_i$, we use the semi empirical formula in terms of mass number $A_i$ as:

$$R_i = 1.28A_i^{1/3} - 0.76 + 0.8A_i^{-1/3} \qquad (7)$$

**2.2. The fusion cross section**

To describe the fusion reactions at energies not too much above the barrier and at higher energies, the barrier penetration model developed by C. Y. Wong [40] has been widely used for the last four decades, which obviously explains the experimental result properly.

Following Thomas [51], Huizenga and Igo [52] and Rasmussen and Sugawara [53], Wong approximated the various barriers for different partial waves by inverted harmonic oscillator potentials of height $E_\ell$ and frequency $\omega_\ell$. For energy $E$, using the probability for the absorption of $\ell^{th}$ partial wave given by Hill-Wheeler formula [54], Wong arrived at the total cross section for the fusion of two nuclei by quantum mechanical penetration of simple one-dimensional potential barrier as:

$$\sigma = \frac{\pi}{k^2} \sum_\ell \frac{2\ell + 1}{1 + \exp[2\pi (E_\ell - E) / \hbar\omega_\ell]} \tag{8}$$

where $k = \sqrt{\frac{2\mu E}{\hbar^2}}$. Here $\hbar\omega_\ell$ is the curvature of the inverted parabola. Using some parameterizations in the region $\ell = 0$ and replacing the sum in Eq. (8) by an integral Wong gave the reaction cross section as:

$$\sigma = \frac{R_0^2 \hbar\omega_0}{2E} \ln\left\{1 + \exp\left[\frac{2\pi (E - E_0)}{\hbar\omega_0}\right]\right\} \tag{9}$$

where $R_0$ is the barrier radius and $E_0$ is the barrier height.

For relatively large values of $E$, the above result reduces to the well-known formula:

$$\sigma = \pi R_0^2 \left[1 - \frac{E_0}{E}\right] \tag{10}$$

## 3. Results and discussions

In the study of cold reaction valleys in the radioactive decay of super heavy nucleus $^{286}$112 [37], in addition to $^{4}$He+$^{282}$110 system, $^{8}$Be+$^{278}$Hs, $^{10}$Be+$^{276}$Hs, $^{14}$C+$^{272}$Sg, $^{20}$O+$^{266}$Rf, $^{24}$Ne+$^{262}$No, $^{26}$Ne+$^{260}$No etc were found to be the possible candidates for binary splittings. Moreover moving on to the fission region, there were two deep regions each consisting of three comparable minima. In the first deep region, due to the double magicity of $^{208}$Pb, a first minimum corresponds to the splitting $^{78}$Zn + $^{208}$Pb, due to the magic neutron shell $N$ = 126 of $^{206}$Hg, a second minimum corresponds to the splitting $^{80}$Ge + $^{206}$Hg and due to the magic neutron shell $N$ = 50 of $^{82}$Ge a third minimum corresponds to the splitting $^{82}$Ge + $^{204}$Hg were observed. For the second deep region, the first two minima involved $^{130}$Sn+$^{156}$Sm and $^{132}$Sn + $^{154}$Sm splittings and the third minimum comes from the splitting $^{134}$Te + $^{152}$Nd, due to the presence of Z = 50 and N = 82 magic shells. Since the above discussed clusters and daughter nuclei lie in the cold valleys, they are the optimal cases of asymmetric/symmetric binary splittings and hence can be identified as the optimal projectile- target combinations for the synthesis of super heavy element, with considerations to the nature of interaction barrier, potential pocket for a realistic depth and nuclear stability.

Taking Coulomb and proximity potential as the scattering potential, we have calculated the interaction barriers for the fusion of all the above identified optimal projectile-target combinations in the case of super heavy $^{286}112$ nucleus, against the distance between the centers of the projectile and target and the corresponding barrier height $E_0$ and the barrier radius $R_0$ noted with $\ell = 0$ and the values are given in Table 1. Moreover, near and above the barrier, the total fusion cross-sections for all the above systems have also been calculated by using the values of $E_B$ and $R_B$ and using Eqs. (8) and (10). It is found that the computed fusion cross sections for all the systems are of the order of several millibarn.

As a systematic study for predicting the most suitable projectile target combination for heavy ion fusion experiment, initially, take the projectile-target combinations $^{134}$Te +$^{152}$Nd, $^{132}$Sn + $^{154}$Sm and $^{130}$Sn+$^{156}$Sm that are found in the second deep region in the cold valleys of $^{286}112$ nucleus. While plotting the interaction barrier against the distance between the centers of the projectile and target for the above three combinations as shown in Figs.1 (a), (b) and (c), it should be noted that for the barrier height $E_B$ is maximum for $^{134}$Te +$^{152}$Nd with minimum barrier radius $R_B$; but the potential pockets that are to be appreciable for the fusion to takes place, are shallow in all the three cases and hence cannot be used as a suitable projectile-target combination for heavy ion fusion reactions. Moreover, the projectiles are comparatively heavy and while noting the half lives, none of the projectiles are stable also.

While analyzing the interaction barriers for the rest of the combinations given in Table. 1, it is observed that the potential pockets are appreciable in the cases of $^{82}$Ge + $^{204}$Hg system onwards, as shown in Fig. 1(d). Now, for a detailed analysis of the fusion possibility in the first deep region in the cold valleys, consider the systems $^{82}$Ge + $^{204}$Hg, $^{80}$Ge + $^{206}$Hg and $^{78}$Zn + $^{208}$Pb. The presence of magic neutron shell N=50 of $^{82}$Ge in the first system, the magic neutron shell N=126 of $^{206}$Hg in the second system and the presence of double magicity of $^{208}$Pb in the third system, along with moderately spanned potential pockets make the systems as suitable projectile-target combinations for the synthesis of super heavy nucleus $^{286}112$, which is in good agreement with the predictions in Ref. [31].

Further, in an attempt to predict more suitable projectile-target combinations from the binary splittings, which are having good potential pockets, we have considered the projectiles and targets having comparatively large half lives and the systems $^{64}Fe+^{222}Rn$, $^{54}Ti+^{232}Th$, $^{50}Ca+^{236}U$, $^{48}Ca+^{238}U$, $^{46}Ar+^{240}Pu$, $^{44}Ar+^{242}Pu$, $^{40}S+^{246}Cm$, $^{38}S+^{248}Cm$ and $^{20}O+^{266}Rf$ are found to be feasible for fusion experiments. For a comparison, the interaction barriers for the eight systems starting from $^{48}Ca+^{238}U$ to $^{32}Mg+^{254}Fm$ are shown in Figs. 2 and 3. Near and above the barrier, the computed cross sections for the above systems using Eq.(8) and (10) are given in Table. 2 and the corresponding excitation functions (σ versus $E_{CM}$ plots) are shown in Figs. 4 and 5. Even though Eq.(10) is an approximation of Eq. (9) at higher energies, the fusion cross sections can be computed very easily with Eq.(10) by noting the values of $E_0$ and $R_0$ from Table. 1. It can be seen that in Table. 2, cross sections computed using Eq. (10) almost matches with the result of Eq. (8), as the energies are above the barrier height. Considering the stability of the nuclei based on half lives, it can be seen that the systems $^{48}Ca+^{238}U$, $^{38}S+^{248}Cm$ and $^{44}Ar+^{242}Pu$ systems are the better optimal projectile–target combinations for heavy ion fusion experiments for the synthesis of super heavy nucleus $^{286}112$. It should be noted that all the targets $^{238}U$, $^{248}Cm$ and $^{242}Pu$ are relatively stable, where as the doubly magic $^{48}Ca$ is the only stable projectile and the half life of $^{38}S$ is 170 minute and $^{44}Ar$ is 11.87 minute.

## 4. Conclusions

We have calculated the interaction barriers for the fusion of all the projectile-target combinations identified in the cold valleys of super heavy $^{286}112$ nucleus, against the distance between the centers of the projectile and target by taking Coulomb and proximity potential as the scattering potential. Near and above the barrier, the total fusion cross-sections for all the systems also have been calculated and it is found that the computed fusion cross sections for all the systems are of the order of several millibarn. The systems $^{82}Ge + ^{204}Hg$, $^{80}Ge + ^{206}Hg$ and $^{78}Zn + ^{208}Pb$ in the deep region of cold valley and the systems $^{64}Fe+^{222}Rn$, $^{54}Ti+^{232}Th$, $^{50}Ca+^{236}U$, $^{48}Ca+^{238}U$, $^{46}Ar+^{240}Pu$, $^{44}Ar+^{242}Pu$, $^{40}S+^{246}Cm$, $^{38}S+^{248}Cm$ and $^{20}O+^{266}Rf$ in the cold valleys are identified as the better projectile-target combinations for the synthesis super heavy nucleus $^{286}112$. While considering the nature of potential pockets and half lives of colliding nuclei, the systems $^{82}Ge + ^{204}Hg$, $^{80}Ge + ^{206}Hg$ and $^{78}Zn + ^{208}Pb$ in the deep cold valley and the systems $^{48}Ca+^{238}U$, $^{38}S+^{248}Cm$ and $^{44}Ar+^{242}Pu$ in the other cold valleys give maximum probability for the synthesis of super heavy nucleus $^{286}112$.

## References

[1] H. Becquerel, Compt. Rend. **122** (1896) 420.

[2] P. Armbruster, Ann. Rev .Nucl. Part. Sci. **35** (1985) 135.

[3] P. Armbruster, Ann. Rev. Nucl. Part. Sci. **50** (2000) 411.

[4] K. P. Santhosh, B. Priyanka, Phys. Rev. **C 87** (2013) 064611.

[5] G. Münzenberg, P. Armbruster, H. Folger, F. P. Heßberger, S. Hofmann, J. Keller, K. Poppensieker, W. Reisdorf, K. H. Schmidt, H. J. Schött, M. E. Leino, R. Hingmann, Z. Phys. **A 317** (1984) 235.

[6] N. Bohr, J. A. Wheeler, Phys. Rev. **56** (1939) 426.

[7] G. Münzenberg, S. Hofmann, F. P. Heßberger, W. Reisdorf, K. H. Schmidt, J. H. R. Schneider, P. Armbruster, C. C. Sahm, B. Thuma, Z. Phys. **A 300** (1981) 107.

[8] Yu. Ts. Oganessian, A. S. Ilinov, A. G. Demin, S. P. Tretyakova, Nucl. Phys. **A 239** (1975) 353

[9] G. Münzenberg, P. Armbruster, F. P. Heßberger, S. Hofmann, K. Poppensieker, W. Reisdorf, J. H. R. Schneider, W. F. W. Schneider, K. H. Schmidt, C. C. Sahm, D. Vermeulen, Z. Phys. **A 309** (1982) 89.

[10] G. Münzenberg, W. Reisdorf, S. Hofmann, Y. K. Agarwal, F. P. Heßberger, K. Poppensieker, J. R. H. Schneider, W. F. W. Schneider, K. H. Schmidt, H. J. Schött, P. Armbruster, C. C. Sahm, D. Vermeulen, Z. Phys. **A 315** (1984) 145.

[11] G. Münzenberg, P. Armbruster, H. Folger, F. P. Heßberger, S. Hofmann, J. Keller, K. Poppensieker, W. Reisdorf, K. H. Schmidt, H. J. Schött, M. Leino, R. Hingmann, Z. Phys. **A 317** (1984) 235.

[12] S. Hofmann, V. Ninov, F. P. Heßberger, P. Armbruster, H. Folger, G. Münzenberg, H. J. Schött, A. G. Popeko, A. V. Yeremin, A. N. Andreyev, S. Saro, R. Janik, M. Leino, Z. Phys. **A 350** (1995) 277.

[13] S. Hofmann, V. Ninov, F. P. Heßberger, P. Armbruster, H. Folger, G. Münzenberg, H. J. Schött, A. G. Popeko, A. V. Yeremin, A. N. Andreyev, S. Saro, R. Janik, M. Leino, Z. Phys. **A 350** (1995) 281.

[14] S. Hofmann, V. Ninov, F. P. Heβberger, P. Armbruster, H. Folger, G. Münzenberg, H. J. Schött, A. G. Popeko, A. V. Yeremin, S. Saro, R. Janik, M. Leino, Z. Phys. **A 354** (1996) 229.

[15] Yu. Ts. Oganessian, V. K. Utyonkov, Yu. V. Lobanov, F. Sh. Abdullin, A. N. Polyakov, R. N. Sagaidak, I. V. Shirokovsky, Yu. S. Tsyganov, A. A. Voinov, G. G. Gulbekian, S. L. Bogomolov, B. N. Gikal, A. N. Mezentsev, V. G. Subbotin, A. M. Sukhov, K. Subotic, V. I. Zagrebaev, G. K. Vostokin, M. G. Itkis, R. A. Henderson, J. M. Kenneally, J. H. Landrum, K. J. Moody, D. A. Shaughnessy, M. A. Stoyer, N. J. Stoyer, P. A. Wilk, Phys. Rev. **C 76** 2007 011601(R)

[16] L. Stavsetra, K. E. Gregorich, J. Dvorak, P. A. Ellison, I. Dragojevic, M. A. Garcia, H. Nitsche, Phys. Rev. Lett. **103** (2009) 132502.

[17] Yu. Ts. Oganessian, V. K. Utyonkov, Yu. V. Lobanov, F. Sh. Abdullin, A. N. Polyakov, I. V. Shirokosvsky, Yu. S. Tsyganov, G. G. Gulbekian, S. L. Bogomolov, A. N. Mezentsev, S. Iliev, V. G. Subbotin, A. M. Sukhov, A. A. Voinov, G. V. Buklanov, K. Subotic, V. I. Zagrebaev, M. G. Itkis, J. B. Patin, K. J. Moody, J. F. Wild, M. A. Stoyer, N. J. Stoyer, D. A. Shaughnessy, J. M. Kenneally, R. W. Lougheed, Phys. Rev. **C 69** (2004) 021601 (R).

[18] Yu. Ts. Oganessian, V. K. Utyonkov, S. N. Dmitriev, Yu. V. Lobanov, M. G. Itkis, A. N. Polyakov, Yu. S. Tsyganov, A. N. Mezentsev, A. V. Yeremin, A. A. Voinov, E. A. Sokol, G. G. Gulbekian, S. L. Bogomolov, S. Iliev, V. G. Subbotin, A. M. Sukhov, G. V. Buklanov, S. V. Shishkin, V. I. Chepygin, G. K. Vostokin, N. V. Aksenov, M. Hussonnois, K. Subotic, V. I. Zagrebaev, K. J. Moody, J. B. Patin, J. F. Wild, M. A. Stoyer, N. J. Stoyer, D. A. Shaughnessy, J. M. Kenneally, P. A. Wilk, R. W. Lougheed, H. W. Gäggeler, D. Schumann, H. Bruchertseifer, R. Eichler, Phys. Rev. **C 72** (2005) 034611.

[19] Yu. Ts. Oganessian, Nucl. Phys. **A 685** (2001) 17c.

[20] Yu. Ts. Oganessian, V. K. Utyonkov, Yu. V. Lobanov, F. Sh. Abdullin, A. N. Polyakov, I. V. Shirokovsky, Yu. S. Tsyganov, G. G. Gulbekian, S. L. Bogomolov, A. N. Mezentsev, S. Iliev, V. G. Subbotin, A. M. Sukhov, A. A. Voinov, G. V. Buklanov, K. Subotic, V. I. Zagrebaev, M. G. Itkis, J. B. Patin, K. J. Moody, J. F. Wild, M. A. Stoyer, N. J. Stoyer, D. A. Shaughnessy, J. M. Kenneally, R. W. Lougheed, Phys. Rev. **C 69** (2004) 021601 (R).

[21] Yu. Ts. Oganessian, V. K. Utyonkov, Yu. V. Lobanov, F. Sh. Abdullin, A. N. Polyakov, I. V. Shirokovsky, Yu. S. Tsyganov, G. G. Gulbekian, S. L. Bogomolov, B. N. Gikal, A. N. Mezentsev, S. Iliev, V. G. Subbotin, A. M. Sukhov, A. A. Voinov, G. V. Buklanov, K. Subotic, V. I. Zagrebaev, M. G. Itkis, J. B. Patin, K. J. Moody, J. F. Wild, M. A. Stoyer,

N. J. Stoyer, D. A. Shaughnessy, J. M. Kenneally, R. W. Lougheed, Nucl. Phys. **A 734** (2004) 109.

[22] Yu. Ts. Oganessian, J. Phys. G: Nucl. Part. Phys **34** (2007) R165.

[23] M. A. Stoyer, Nature **442** (2006) 876.

[24] S. Hofmann, G. Münzenberg, Rev. Mod. Phys. **72** (2000) 733.

[25] Yu. Ts. Oganessian, F. Sh. Abdullin, P. D. Bailey, D. E. Benker, M. E. Bennett, S. N. Dmitriev, J. G. Ezold, J. H. Hamilton, R. A. Henderson, M. G. Itkis, Yu. V. Lobanov, A. N. Mezentsev, K. J. Moody, S. L. Nelson, A. N. Polyakov, C. E. Porter, A. V. Ramayya, F. D. Riley, J. B. Roberto, M. A. Ryabinin, K. P. Rykaczewski, R. N. Sagaidak, D. A. Shaughnessy, I. V. Shirokovsky, M. A. Stoyer, V. G. Subbotin, R. Sudowe, A. M. Sukhov, Yu. S. Tsyganov, V. K. Utyonkov, A. A. Voinov, G. K. Vostokin, P. A. Wilk , Phys. Rev. Lett. **104** (2010) 142502.

[26] Yu. Ts. Oganessian, F. Sh. Abdullin, P. D. Bailey, D. E. Benker, M. E. Bennett, S. N. Dmitriev, J. G. Ezold, J. H. Hamilton, R. A. Henderson, M. G. Itkis, Yu. V. Lobanov, A. N. Mezentsev, K. J. Moody, S. L. Nelson, A. N. Polyakov, C. E. Porter, A. V. Ramayya, F. D. Riley, J. B. Roberto, M. A. Ryabinin, K. P. Rykaczewski, R. N. Sagaidak, D. A. Shaughnessy, I. V. Shirokovsky, M. A. Stoyer, V. G. Subbotin, R. Sudowe, A. M. Sukhov, R. Taylor, Yu. S. Tsyganov, V. K. Utyonkov, A. A. Voinov, G. K. Vostokin, P. A. Wilk, Phys. Rev. **C 83** (2011) 054315.

[27] A. Sandulescu, D. N. Poenaru, W. Greiner, Fiz. Elem. Chastits At. Yadra **11** (1980) 1334; Sov. J. Part. Nucl. **11** (1980) 528.

[28] R. K. Gupta, Fiz. Elem. Chastits At. Yadra **8 (**1977) 717; Sov. J. Part. Nucl. **8** (1977) 289.

[29] J. A. Maruhn, W. Greiner, W. Scheid, Heavy-Ion Collisions (North-Holland, Amsterdam) Vol. 2 p. (1980) 399.

[30] D. N. Poenaru, W. Greiner, R. Gherghescu, Phys. Rev. **C 47 (**1993) 2030.

[31] R. K. Gupta, J. Phys. G: Nucl. Part. Phys. **21** (1995) L89.

[32] H. J. Rose, G. A. Jones, Nature **307** (1984) 245.

[33] Yu. Ts. Oganessian, V. L. Mikheev, S. P. Tretyakova, Report No. E7 (1993) 93-57 JINR (Dubna)

[34] Yu. Ts. Oganessian, Yu. A. Lazarev, V. L. Mikheev, Yu. A. Muzychka, I. V. Shirokovsky, S. P. Tretyakova, V. K. Utyonkov, Z. Phys. **A 349** (1994) 341.

[35] A. Guglielmetti, B. Blank, R. Bonetti, Z. Janas, H. Keller, R. Kirchner, O. Klepper, A. Piechaczek, A. Plochocki, G. Poli, P. B. Price, E. Roeckl, K. Schmidt, J. Szerypo, A. J. Westphal, Nucl. Phys. **A 583** (1995) 867c

[36] A. Guglielmetti, R. Bonetti, G. Poli, P. B. Price, A. J. Westphal, Z. Janas, H. Keller, R. Kirchner, O. Klepper, A. Piechaczek, E. Roeckl, K. Schmidt, A. Plochocki, J. Szerypo, B. Blank, Phys. Rev. **C** 52 (1995) 740.

[37] K. P. Santhosh, S. Sabina, Phys. At. Nucl. **75** (2012) 973.

[38] K. P. Santhosh, A. Joseph, Pramana J. Phys. **58** (2002) 611.

[39] M. G. Itkis *et al.*, 2001 Proceedings of the International Workshop on Fusion Dynamics at the Extremes, Dubna, 2000, Ed. by Yu. Ts. Oganessian and V. I. Zagrebaev (*World Sci., Singapore*) p. 93

[40] C. Y. Wong, Phys. Rev. Lett. **31** (1973) 766.

[41] N. Rowley, G. R. Satchler, P. H. Stelson, Phys. Lett. **B 254** (1991) 25.

[42] R. K. Puri, R. K. Gupta, Phys. Rev. **C 45** (1992) 1837.

[43] J. R. Leigh, M. Dasgupta, D. J. Hinde, J. C. Mein, C. R. Morton, R. C. Lemmon, J. P. Lestone, J. O. Newton, H. Timmers, J. X. Wei, N. Rowley, Phys. Rev. **C 52** (1995) 3151.

[44] H. Timmers, J. R. Leigh, N. Rowley, A. M. Stefanini, D. Ackermann, S. Beghini, L. Corradi, M. Dasgupta, J. H. He, D. J. Hinde, J. C. Mein, G. Montagnoli, C. R. Morton, J. O. Newton, F. Scarlassara, G. F. Segato, J. Phys. G: Nucl. Part. Phys. **2**3 (1997) 1175.

[45] K. Hagino, N. Rowley, A. T. Kruppa, Comput. Phys. Commun.**123** (1999) 143.

[46] J. O. Newton, R. D. Butt, M. Dasgupta, D. J. Hinde, I. I. Gontchar, C. R. Morton, K. Hagino, Phys. Rev. **C 70** (2004) 024605.

[47] P. R. S. Gomes, I. Padron, E. Crema, O. A. Capurro, Fernandez Niello, A. Arazi, G. V. Marti, J. Lubian, M. Trotta, A. J. Pacheco, J. E. Testoni, M. D. Rodriguez, M. E. Ortega, L. C. Chamon, R. M. Anjos, R. Veiga, M. Dasgupta, D. J. Hinde, K. Hagino Phys. Rev. **C 73** (2006) 064606.

[48] N. Wang, Zhuxia Li, W. Scheid, J. Phys. G: Nucl. Part. Phys. **34** (2007) 1935.

[49] K. P. Santhosh, V. Bobby Jose, A. Joseph, K. M. Varier, Nucl. Phys. **A 817** (2009) 35.

[50] J. Blocki, J. Randrup, W. J. Swiatecki, C. F. Tsang, Ann. Phys. (N.Y) **105** (1977) 427.

[51] T. D. Thomas, Phys. Rev. **116** (1959) 703.

[52] J. R. Huizenga, G. Igo, Nucl. Phys. **29** (1962) 462.

[53] J. O. Rasmussen, K. Sugawara- Tanabe, Nucl. Phys. **A 171** (1971) 497.

[54] D. L. Hill, J. A. Wheeler, Phys. Rev. **89** (1953) 1102.

**Table 1.** Barrier height and barrier radius for the systems in the cold valleys of $^{286}112$ nuclei using Coulomb and proximity potentials.

| Reaction | Barrier height $E_0$ (MeV) | Barrier radius $R_0$ (fm) |
|---|---|---|
| $^{134}Te+^{152}Nd$ | 312.400 | 13.200 |
| $^{132}Sn+^{154}Sm$ | 310.311 | 13.248 |
| $^{130}Sn+^{156}Sm$ | 310.413 | 13.244 |
| $^{82}Ge+^{204}Hg$ | 259.089 | 13.279 |
| $^{80}Ge+^{206}Hg$ | 259.585 | 13.209 |
| $^{78}Zn+^{208}Pb$ | 249.235 | 13.288 |
| $^{68}Ni+^{218}Po$ | 240.253 | 13.170 |
| $^{66}Fe+^{220}Rn$ | 228.196 | 13.243 |
| $^{64}Fe+^{222}Rn$ | 228.806 | 13.166 |
| $^{62}Cr+^{224}Ra$ | 215.900 | 13.237 |
| $^{60}Cr+^{226}Ra$ | 216.520 | 13.207 |
| $^{58}Cr+^{228}Ra$ | 217.131 | 13.225 |
| $^{56}Ti+^{230}Th$ | 203.392 | 13.192 |
| $^{54}Ti+^{232}Th$ | 204.059 | 13.108 |
| $^{52}Ca+^{234}U$ | 189.405 | 13.173 |
| $^{50}Ca+^{236}U$ | 190.080 | 13.135 |
| $^{48}Ca+^{238}U$ | 190.796 | 13.047 |
| $^{46}Ar+^{240}Pu$ | 175.226 | 13.106 |
| $^{44}Ar+^{242}Pu$ | 175.945 | 13.063 |
| $^{42}S+^{244}Cm$ | 159.448 | 13.019 |
| $^{40}S+^{246}Cm$ | 160.200 | 13.022 |
| $^{38}S+^{248}Cm$ | 160.966 | 12.974 |
| $^{34}Si+^{252}Cf$ | 144.301 | 12.918 |
| $^{32}Mg+^{254}Fm$ | 125.925 | 12.969 |
| $^{28}Mg+^{258}Fm$ | 127.495 | 12.788 |
| $^{26}Ne+^{260}No$ | 108.069 | 12.820 |
| $^{24}Ne+^{262}No$ | 108.868 | 12.748 |
| $^{20}O+^{266}Rf$ | 89.200 | 12.689 |
| $^{14}C+^{272}Sg$ | 69.254 | 12.447 |
| $^{10}Be+^{276}Hs$ | 47.276 | 12.399 |
| $^{8}Be+^{278}Hs$ | 48.150 | 12.150 |
| $^{4}He+^{282}110$ | 51.002 | 10.865 |

**Table 2.** Computed fusion cross sections for the system $^{48}Ca+^{238}U$ to $^{32}Mg+^{254}Fm$ systems.

| Reaction | $E_{CM}$ (MeV) | σ (mb) | | Reaction | $E_{CM}$ (MeV) | σ (mb) | |
|---|---|---|---|---|---|---|---|
| | | Eq.(8) | Eq.(10) | | | Eq.(8) | Eq.(10) |
| $^{48}Ca+^{238}U$ | 192 | 34.00 | 33.53 | $^{40}S+^{246}Cm$ | 162 | 61.87 | 59.20 |
| | 194 | 86.32 | 88.32 | | 164 | 125.83 | 123.45 |
| | 196 | 139.98 | 141.98 | | 166 | 182.26 | 186.15 |
| | 198 | 197.92 | 194.57 | | 168 | 249.16 | 247.32 |
| | 200 | 246.87 | 246.10 | | 170 | 301.41 | 307.14 |
| | 202 | 290.89 | 296.66 | | 172 | 357.95 | 365.52 |
| $^{46}Ar+^{240}Pu$ | 176 | 24.56 | 23.73 | $^{38}S+^{248}Cm$ | 162 | 35.29 | 33.75 |
| | 178 | 85.39 | 84.10 | | 164 | 101.46 | 97.83 |
| | 180 | 142.72 | 143.12 | | 166 | 163.17 | 160.36 |
| | 182 | 196.37 | 200.85 | | 168 | 217.79 | 221.41 |
| | 184 | 257.84 | 257.33 | | 170 | 279.45 | 281.01 |
| | 186 | 305.47 | 312.59 | | 172 | 335.44 | 339.24 |
| $^{44}Ar+^{242}Pu$ | 178 | 61.48 | 61.89 | $^{34}Si+^{252}Cf$ | 146 | 59.62 | 61.01 |
| | 180 | 119.17 | 120.78 | | 148 | 132.34 | 131.10 |
| | 182 | 177.90 | 178.37 | | 150 | 198.62 | 199.20 |
| | 184 | 228.51 | 234.7 | | 152 | 264.74 | 265.37 |
| | 186 | 284.80 | 289.84 | | 154 | 325.55 | 330.21 |
| | 188 | 335.33 | 343.79 | | 156 | 391.87 | 393.20 |
| $^{42}S+^{244}Cm$ | 160 | 16.37 | 18.37 | $^{32}Mg+^{254}Fm$ | 126 | 4.55 | 3.18 |
| | 162 | 81.87 | 83.88 | | 128 | 86.74 | 85.60 |
| | 164 | 151.88 | 147.80 | | 130 | 169.58 | 165.49 |
| | 166 | 212.20 | 210.18 | | 132 | 237.92 | 242.93 |
| | 168 | 270.07 | 271.08 | | 134 | 316.55 | 318.11 |
| | 170 | 335.65 | 330.54 | | 136 | 388.65 | 391.05 |

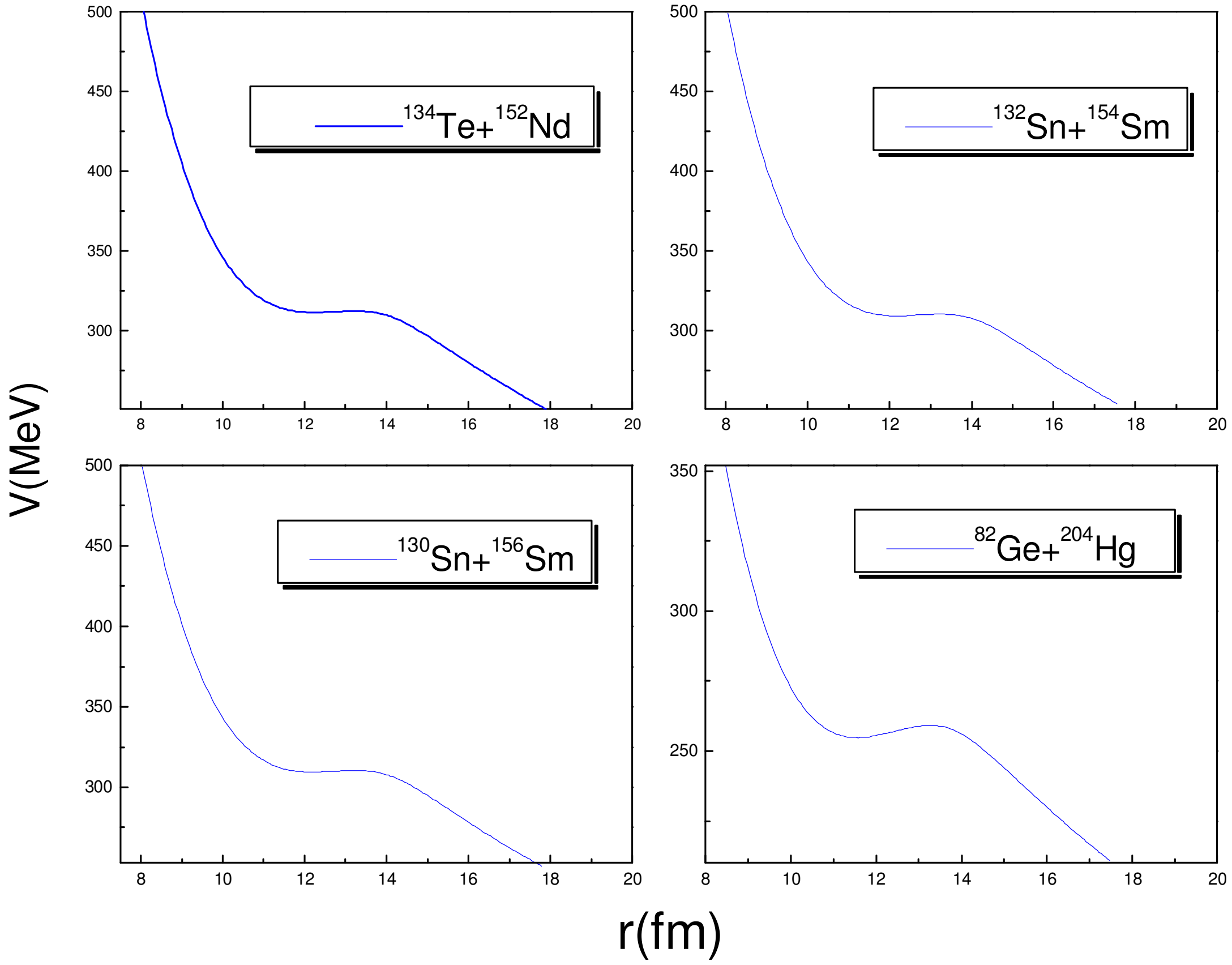

**Fig. 1.** Scattering potential for the reactions of $^{134}$Te +$^{152}$Nd, $^{132}$Sn + $^{154}$Sm, $^{130}$Sn+$^{156}$Sm and $^{82}$Ge+$^{204}$Hg, systems consisting of repulsive Coulomb and centrifugal potentials and attractive nuclear proximity potential.

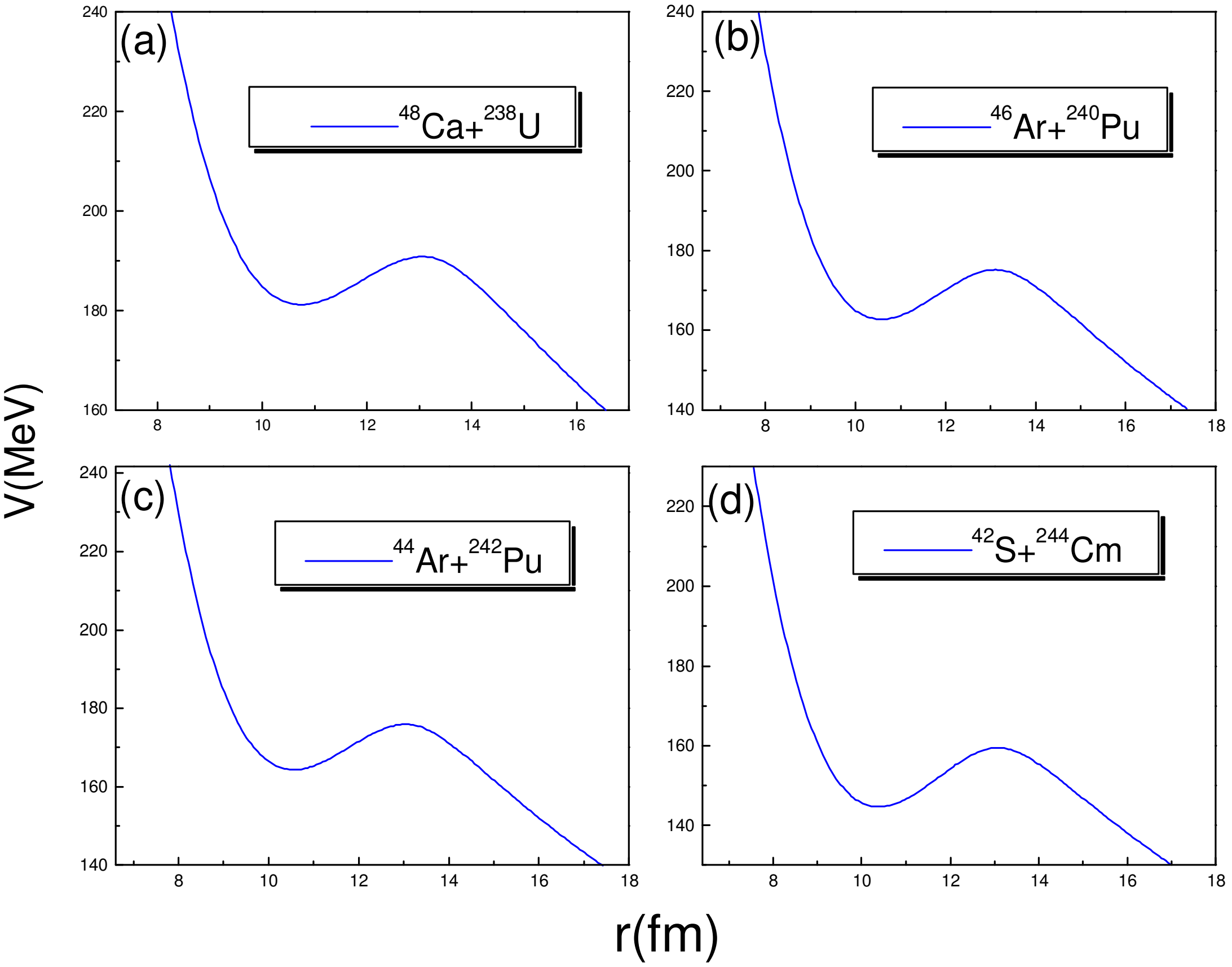

**Fig. 2.** Scattering potential for the reactions of $^{48}$Ca+$^{238}$U, $^{46}$Ar+$^{240}$Pu, $^{44}$Ar+$^{242}$Pu and $^{42}$S+$^{244}$Cm systems.

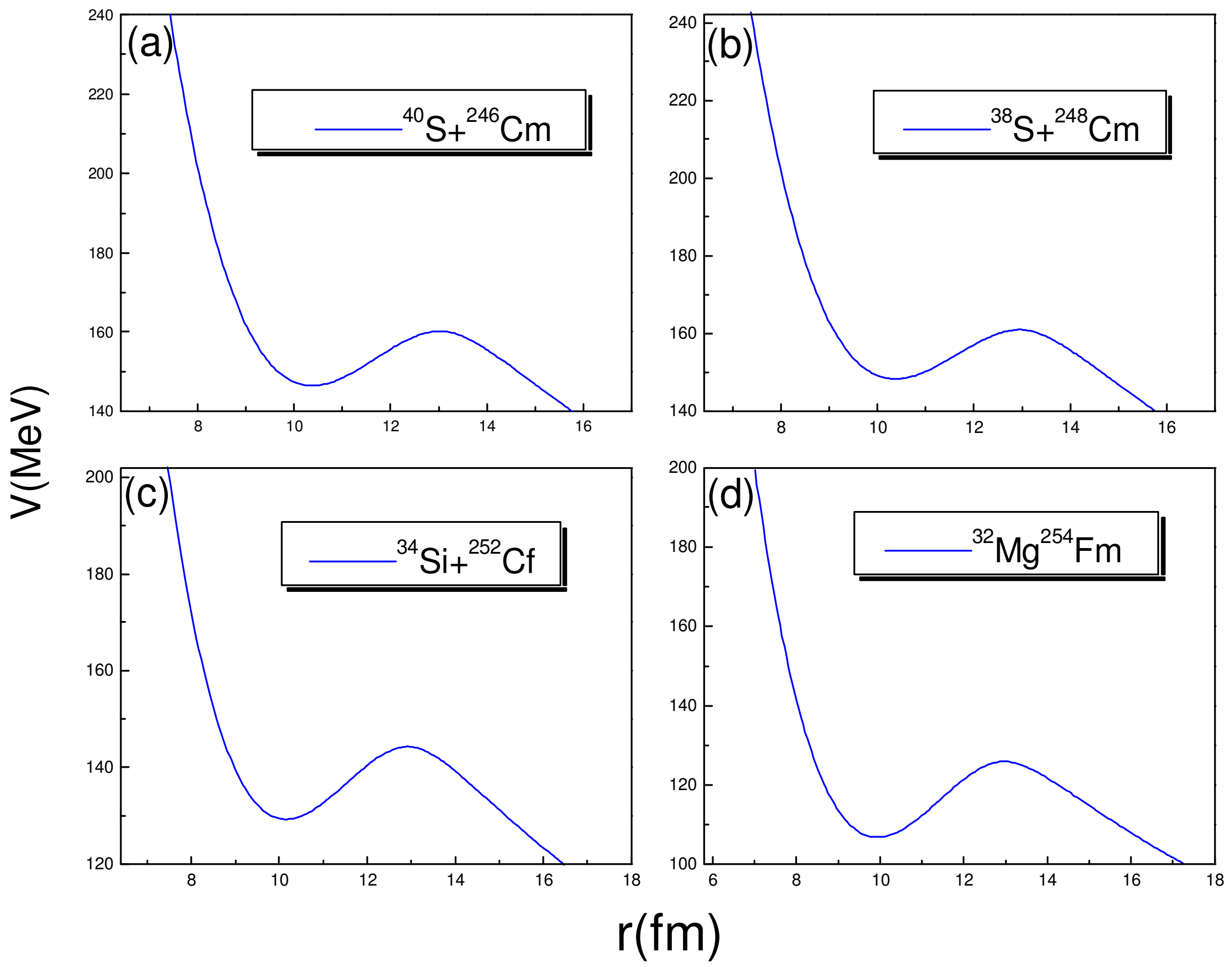

**Fig. 3.** Scattering potential for the reactions of $^{40}$S+$^{246}$Cm, $^{38}$S+$^{248}$Cm, $^{34}$Si+$^{252}$Cf and $^{32}$Mg+$^{254}$Fm systems.

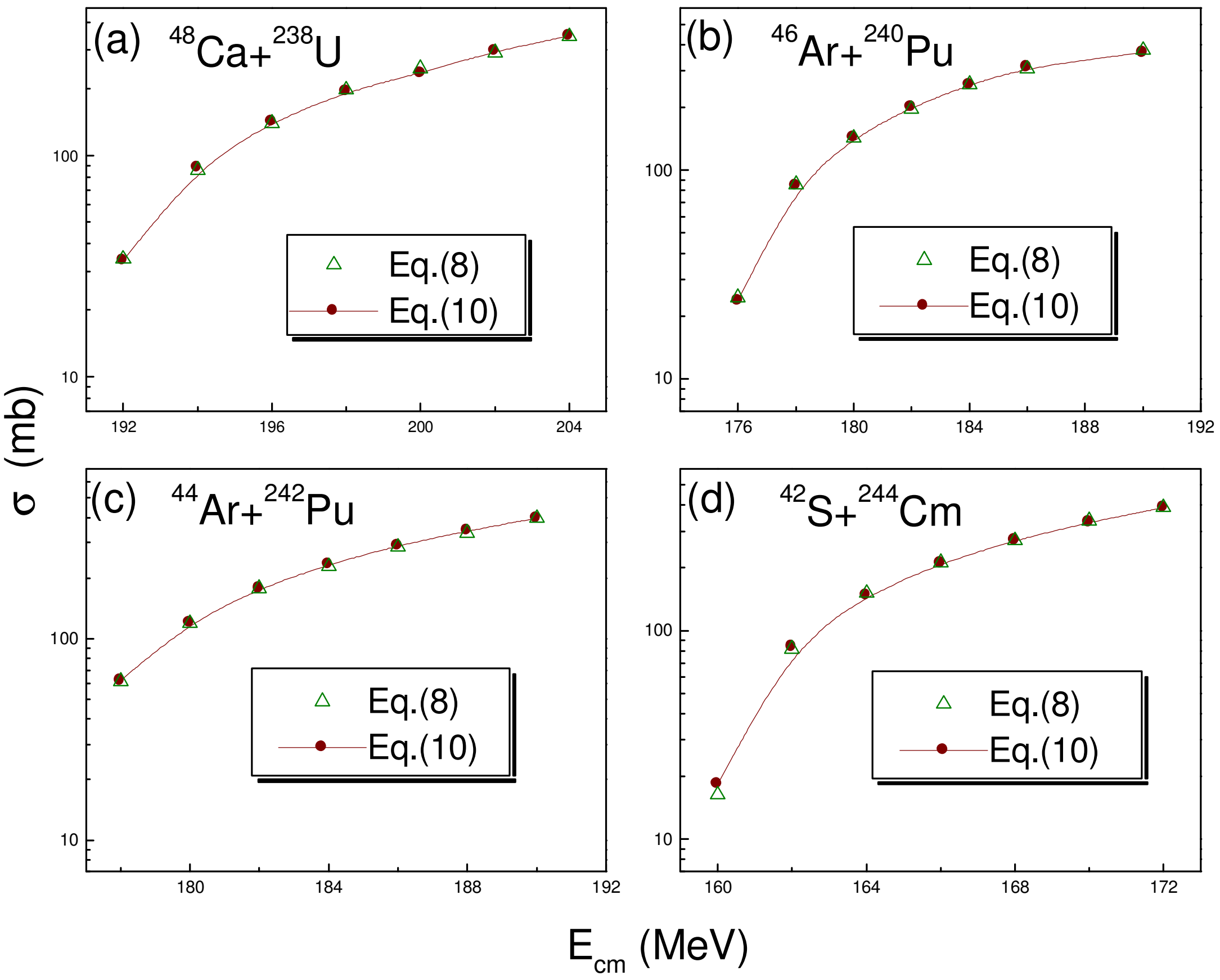

**Fig. 4.** Computed fusion cross sections for the reactions of $^{48}$Ca+$^{238}$U, $^{46}$Ar+$^{240}$Pu, $^{44}$Ar+$^{242}$Pu and $^{42}$S+$^{244}$Cm systems.

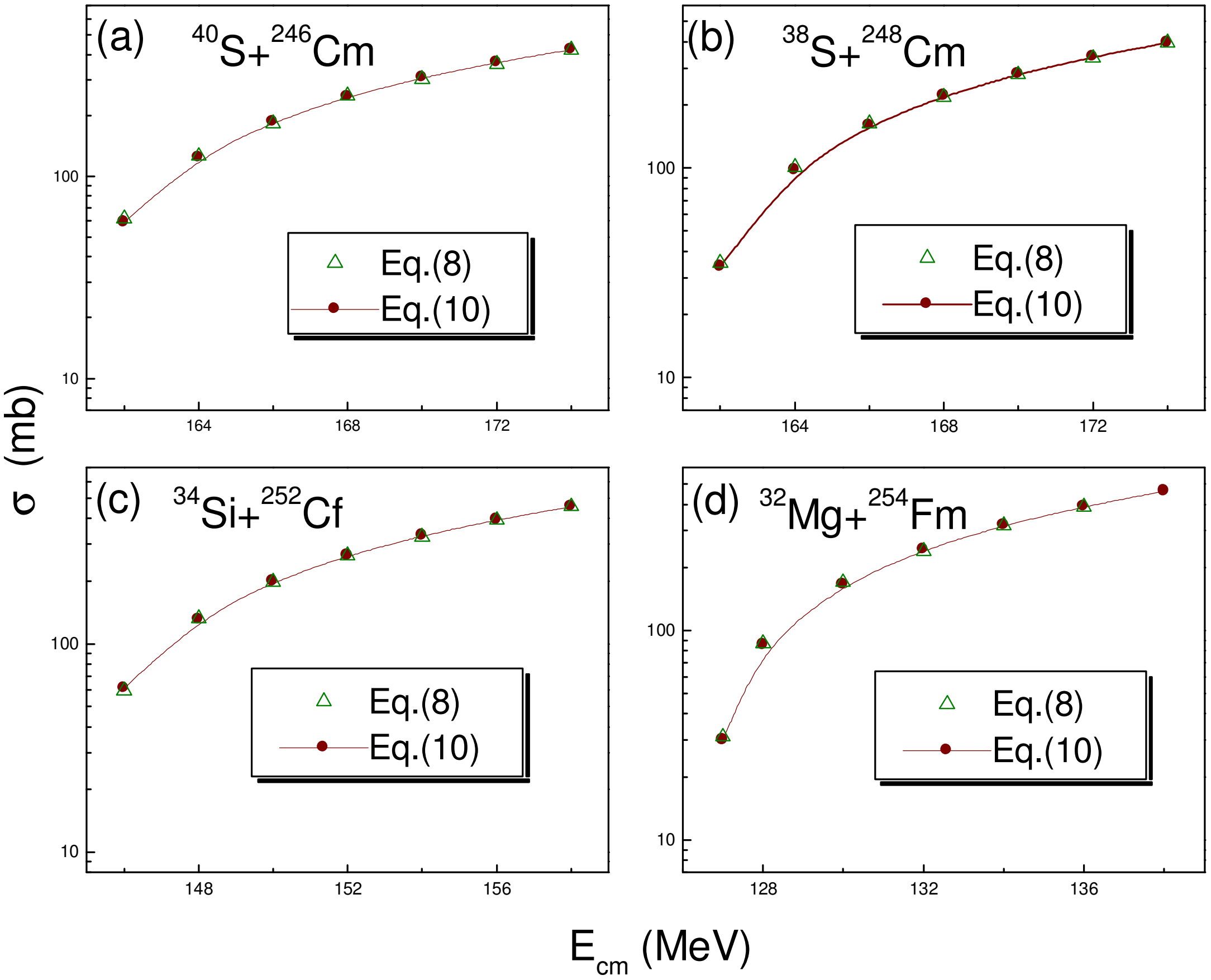

**Fig. 5.** Computed fusion cross sections for the reactions of $^{40}S+^{246}Cm$, $^{38}S+^{248}Cm$, $^{34}Si+^{252}Cf$ and $^{32}Mg+^{254}Fm$ systems